\documentstyle[12pt]{article}

\begin{document}

\begin{titlepage}
\null\vspace{-62pt}

\pagestyle{empty}
\begin{center}

\vspace{1.0truein} {\Large\bf Effects of the cosmological expansion
on the bubble nucleation rate for relativistic first-order phase
transitions}

\vspace{1in}
{\large\bf Dimitrios Metaxas} \\
\vskip .4in
{\it Department of Physics,\\
National Technical University of Athens,\\
Zografou Campus, 15780 Athens, Greece\\
metaxas@central.ntua.gr}\\

\vspace{0.5in}

\vspace{.5in}
\centerline{\bf Abstract}

\baselineskip 18pt
\end{center}

I calculate the first corrections to the dynamical pre-exponential
factor of the bubble nucleation rate for a relativistic first-order
phase transition in an expanding cosmological background by
estimating the effects of the Hubble expansion rate on the critical
bubbles of Langer's statistical theory of metastability. I also
comment on possible applications and problems that arise when one
considers the field theoretical extensions of these results (the
Coleman-De~Luccia and Hawking-Moss instantons and decay rates).

\end{titlepage}

\newpage
\pagestyle{plain}
\setcounter{page}{1}
\newpage

\section{Introduction}

The modern nucleation theory of first-order phase transitions is
based on the results of Langer in \cite{langer} and associated
works. These were generalized to quantum field theory
\cite{coleman}, at finite temperature \cite{linde}, and in curved
spacetime \cite{cdl, hm}. The importance of these considerations is
highlighted by the fact that the original inflationary models based
on a first-order cosmological phase transition \cite{gt, kazanas,
sato, guth} were soon ruled out \cite{gw} and replaced by various
slow-roll models with numerous fine-tuning problems. Of related
recent interest are theories of the landscape and the multiverse
\cite{pol, car} for which a detailed knowledge of the false vacuum
decay rate in curved space-time is of vital importance and may lead
to a cosmological determination of physical parameters such as the
cosmological constant.

Langer's original theory led to the statistical determination of the
bubble nucleation rate, $\Gamma$, that gives the number of critical
sized metastable bubbles of the new phase nucleated per unit volume
and per unit time,
\begin{equation}
\Gamma= \frac{dN}{d^3 x \,dt}=\Omega \,\frac{\kappa}{2\pi}
\exp{(-F/T)}, \label{basic}
\end{equation}
where $\Omega$ and $\kappa$ are kinematical and dynamical factors
respectively. $\Omega$ is proportional to the physical volume of the
system, $\kappa$ is the growth rate of the metastable configuration
and $F$ its free energy. If $\mu$ is a characteristic mass scale of
the theory, of the order of the temperature $T$, then $\Omega$ and
$\kappa$ are of order $\mu^3$ and $\mu$ respectively, giving
$\Gamma$ the correct overall dimensionality.

In the case of a quantum field theory involving a scalar field
$\phi$ and a Euclidean action functional
\begin{equation}
S(\phi)=\int d^4 x \left[\frac{1}{2} (\partial \phi)^2 +
U(\phi)\right],
\end{equation}
with a potential $U(\phi)$ that has a relative minimum (false
vacuum) and an absolute minimum (true vacuum), the bubble nucleation
rate or false vacuum decay rate when the field is trapped in the
false vacuum is given by \cite{coleman}
\begin{equation}
\Gamma=\frac{dN}{d^3 x\,dt}= A\,\exp(-B)
\label{coleman}
\end{equation}
where $B=S_0$ is the Euclidean action of the instanton that is the
solution to the Euclidean equations of motion and $A$ is a
pre-exponential factor that is given by a functional determinant
ratio. If $\mu$ is again a characteristic mass scale of the theory
and $\lambda$ the coupling constant, then $S_0$ is of order
$1/\lambda$ and $A$ of order $\mu^4$.

If the quantum field theory is considered at finite temperature $T$
that is much higher than the inverse critical bubble radius $1/R_*$
(generically $R_*$ is of order $1/\lambda\mu$) then $B=S_3/T$, where
$S_3$ is the three dimensional action of the dimensionally reduced
instanton and $A$ is of order $T^4$ \cite{linde}.

The question that I would like to address here is what happens to
the prefactor $A$ when gravitational effects are taken into account,
as is the case in cosmological applications. It is generally true in
flat space-time, and also when gravitational effects are weak, that
the quantity $B$ gives an exponentially smaller factor, hence is
more important quantitatively. One can, however, very well envision
a situation, in a landscape or multiverse scenario, where both $A$
and $B$, although still close to their flat space values, have an
intricate dependence on the model parameters such as the
cosmological constant, with the net result that the rate $\Gamma$
has an actual maximum at the observed values. In fact, it is not
$\Gamma$ $per$ $se$
 that is expected to have a peak; one is rather interested
in suitably defined quantities that measure the rate of conversion
of physical volume to the new phase \cite{gw, ww} like, for example,
\begin{equation}
p(t)=\exp\left[ -\int_{t_0}^{t} dt' \,\Gamma(t')\frac{4\pi}{3}
          \left(\frac{R_*(t')}{R(t')}+\int_{t'}^{t} dt''\frac{V(t'')}{R(t'')}\right)^3 \right],
\label{probability}
\end{equation}
which gives the probability that an arbitrary point in space remains
in the false vacuum at time $t$. Here $t_0$ is the time that
signifies the onset of the cosmological phase transition, $R(t)$ is
the cosmological scale factor, $R_*(t)$ is the critical bubble
radius at the time of nucleation and $V(t)$ is the velocity with
which the bubble wall is expanding. In any case $\Gamma$ is one of
the main inputs necessary for the calculation of these quantities.

The basic problem for the calculation of gravitational effects on
$A$ is, of course, the lack of a consistent definition of quantum
gravity, so, apart from dimensional considerations, very few methods
have been proposed. In \cite{garriga, basu} the prefactor
was calculated for processes involving the creation of topological 
defects, and in \cite{myself} corrections to $A$ were estimated with the 
use of the renormalization group.
 Here I would like to initiate another
possibility which, although limited in scope from the outset, may be
quite useful and is in fact, in principle, important, namely the
generalization of Langer's original work to the case of curved
space-time. Given that the results of \cite{coleman} and
\cite{linde} are the field theoretical generalizations of the
statistical theory of \cite{langer}, it is notable that the
corresponding theory, of which the work of \cite{cdl, hm} is
supposed to be a generalization, does not exist. The precise
formulation of such a theory would involve the generalization of
thermodynamics and kinetic theory in curved space-time and will be
the subject of future work. Here, instead, I will consider the first
gravitational corrections to the pre-exponential factor for the
previously given description of bubble nucleation in first-order
phase transitions as given in \cite{ck, ck2}. That is, I will be
working in an approximation where the critical bubble radius is much
smaller than the horizon size and neglecting the gravitational
back-reaction, essentially assuming that the related Hawking
temperature is much smaller than the temperature and mass scales of
the system. As expected, the leading effect of the cosmological
expansion will be to increase the bubble nucleation rate and an
estimate of the corrections is given. One may compare this with works
that investigate the stability of classical and semiclassical configurations
in an expanding universe \cite{cooperstock, copeland, metaxas} where similar results 
where found, the main difference here being that the bubble configuration 
considered is already metastable.

 Although the result cannot be
straightforwardly extended to the case of quantum gravity the
corrections derived suggest that a more complete treatment of the
nucleation rate is needed; it is important in first-order
inflationary models, landscape and multiverse scenarios, and may
also be of relevance in various cases of late time and other
first-order cosmological phase transitions \cite{apr, fin1, a1,
osc1, osc2, dej, k1, k2, k3, k4}.

In Sec.~2, which is essentially the second part of the introduction,
I describe the results of \cite{cdl, hm}, the problems and the
relative corrections expected. In Sec.~3 I calculate the first
gravitational corrections to the critical bubble configurations of
Langer's statistical theory of metastability as described in
\cite{ck} and I discuss the results and the approximations involved.
 In Sec.~4 I conclude with some
comments.

\section{Gravitational effects on vacuum decay}

One is interested in a quantum field theory of a scalar field $\phi$
in curved space-time with a metric tensor $g_{\mu\nu}$ and Euclidean
action
\begin{equation}
S=\int \sqrt{-g}\left(\frac{1}{16\pi G}{\cal R}
\,+\,\frac{1}{2}g_{\mu\nu}\partial^{\mu}\phi\,\partial^{\nu}\phi
\,+\,U(\phi)\right)
\end{equation}
where ${\cal R}$ is the Ricci curvature, $g=\det g_{\mu\nu}$, and
for definiteness I will consider a potential $U(\phi)$ that is
everywhere positive and has two minima, as before, a relative (false
vacuum) and an absolute (true vacuum). I will assume again that the
mass scale is $\mu$ and the coupling is $\lambda$. If the field is
trapped in the false vacuum and the value of the potential there is
$\varepsilon$, this will effectively be a cosmological constant
$\Lambda=16\pi G\varepsilon$ and space-time will be de~Sitter with a
Hubble expansion rate $H^2=8\pi G\varepsilon/3$. From this point on
one makes several assumptions using the insight from the flat
space-time results described in the previous section. First of all
one assumes that the solutions to the Euclidean equations of motion
will again describe tunneling and determine the exponential factor,
$B$, that is one expects that Euclidean quantum gravity and the
associated path integral have various features similar to the flat
case.

Then $\Gamma$ will determine the  nucleation rate of bubbles of the
true vacuum, which is also de~Sitter space-time, with a smaller
value of the cosmological constant. In terms of the flat case
critical bubble radius $R_0$, which is of order $1/\mu\lambda$, and
the horizon radius $R_{dS}=1/H$, the results of \cite{cdl} 
estimate
$B$ in the thin-wall approximation as a correction to the flat case value, $B_0$,
\begin{equation}
B=\frac{B_0}{ [1+(R_0/2 R_{dS})^2]^2}
 \label{grtunn}
\end{equation}
and the radius of the bubble in the presence of gravity, $R_*$, as
\begin{equation}
R_*=\frac{R_0}{ [1+(R_0/2 R_{dS})^2]}
\label{deltaR}
\end{equation}
where $R_0 = 3\sigma/\varepsilon$ is the bubble radius in the absence of gravity
and $\sigma$ is the  bubble surface tension which is the same
for the bubble with gravity in the thin wall approximation.

It is implicit in the CDL formalism that some sort of dilute
instanton gas approximation must exist in de~Sitter space-time in
order for the instanton action to exponentiate, that is one expects
an approximation of the sort $R_* << R_{dS}$, in which case the
gravitational corrections in (\ref{grtunn}) do not give an
exponentially small correction to $\Gamma$, in fact they may be
comparable to gravitational corrections that exist in the
pre-exponential factor. The expression for $B$ in the opposite
limiting case where gravitational effects are important, when, for
example, the bubble radius is comparable to the horizon size, may
also be completely different than the Coleman-De~Luccia (CDL)
result, such as is the case in the critical Hawking-Moss (HM)
solution \cite{hm}.  Thus the CDL expression has problems in its
interpretation in both limiting cases. Generally, from a strict
point of view, in the weak gravity limit where the gravitational
corrections in the exponent are of the same order as the
pre-exponential factor, which has not been calculated, one may say
that the CDL expression, although natural and reducible to the flat
case result, is not complete.

Another curious coincidence arises when one ponders the possibility
of a thermal interpretation of the gravitational effects, namely
when one considers the fact that de~Sitter space-time has a
naturally defined Hawking temperature,
\begin{equation}
T_{dS}=\frac{H}{2\pi}.
\end{equation}
Provided that a suitable frame is chosen, and a thermal
interpretation can be given \cite{ejw, batra}, one may expect that,
similarly to the transition from $A\sim\mu^4$ to $A\sim T^4$ in the
limit of high temperature, $T>>1/R_*$, in the flat case, a similar
limit may apply in the gravitational case. This would lead to the
approximation $A\sim H^4$ which is expected to hold in other cases
in the literature \cite{garriga, basu}. However, the translation of
the high temperature approximation in the CDL case would read $H
>>\lambda\mu$, and it is well known \cite{al, dem} that when relations like
$H^2
> 4 \,U''$ hold, the CDL instanton does not even exist. One then may
expect that tunneling is described by the HM solution with the
pre-exponential factor being approximated by $H^4$, again without
much justification.

In view of the possible applications of these results (cosmological
phase transitions, landscape and multiverse scenarios) one would
like to have better descriptions and quantitative estimates for
$\Gamma$. In principle one expects that
\begin{equation}
\Gamma =\frac{dN}{d^3 x\,dt} = \sqrt{-g} \,A(\mu,\lambda,{\cal R},
\Lambda)\,\exp(-B)
 \label{general}
\end{equation}
is the general covariant expression for the production of bubbles of
the true vacuum per unit coordinate four-volume, where $g=\det
g_{\mu\nu}$, $A$ has dimensionality $\mu^4$ and is a function of the
mass scale, $\mu$, of the theory, the coupling constants, $\lambda$,
the Ricci scalar, ${\cal R}$, and the cosmological constant,
$\Lambda$. Also in this expression, $B$ is a dimensionless function
of the same parameters, and the only constraint for both $A$ and $B$
is that they reduce to the flat space-time values in the limit of
weak gravity. One would expect a relation such as (\ref{general}) to
emerge from a saddle point evaluation of a path integral and to be
meaningful for the cases where the bubble radius is much smaller
than the horizon; when they are comparable even an interpretation of
(\ref{general}) is not straightforward.

It should be noted that gravitational corrections to the
pre-exponential factor are expected to exist and can be also
estimated in a similar approximation with an entirely different
method that applies the renormalization group to the CDL expression
\cite{myself}.
 As was mentioned in the introduction, here I will try to obtain some
more insight into this expression by calculating the first
gravitational corrections to the dynamical factor $\kappa$ (of
dimension $\mu$) that appears in (\ref{basic}), as it was calculated
in \cite{ck}. The kinematical prefactor $\Omega$ in
(\ref{basic}) gives an expression of the form $\sqrt{-g}\,\mu^3$
(times again a dimensionless factor that contains gravitational
corrections), and similar corrections apply in the generalization
of the free energy of a metastable system in curved space-time. It is assumed that 
a thermodynamical description of the problem can be given in the rest frame 
of the fluid and the cosmological expansion can be treated as a perturbation.
This is expected to hold provided that a characteristic reaction time, $\tau$,
for the internal fluid interactions is much smaller than $H^{-1}$.
I will also assume that we are in the limits of the thin wall approximation
in order to treat $H$ as constant in the calculation, and also to compare 
with the CDL results. In principle, however, the equations can be solved
self-consistently with a variable $H$.

In summary, the approximations that will be assumed throughout will be that the
bubble radius is much smaller than the horizon, that one is within the limits of the thin wall
approximation, and that relations like $\tau << H^{-1}$ hold. Naturally also the 
temperature and mass scales of the theory are assumed much smaller than the
Planck scale.

The physical situation considered is a metastable relativistic fluid
in a spatial extend that is large enough in order to feel the
cosmological expansion, yet small enough in order to  treat it as a
perturbation. As such, the results are not straightforwardly
generalized to quantum field theory, they are intended, however, to
provide some insight to similar considerations. 
It should be
stressed, also, that what is implied is not a breakdown of the
semiclassical approximations leading to the CDL result, but an
interesting dependence of the pre-exponential factor on the
cosmological expansion rate. This is expected for very small values
of the cosmological constant $\Lambda$, or $R_* << R_{dS}$. It is
suggested that the dependence of $\Gamma$ on $\Lambda$ (or $H$) is
much more intricate than what is described by (\ref{grtunn}) and the
associated expressions of \cite{parke}.

\section{First gravitational corrections to Langer's theory of metastability}

I will calculate the effects of cosmological expansion described by
a metric
\begin{equation}
ds^2 = - dt^2 \, +\, R^2(t) \,[dx^2 +dy^2 + dz^2]
\end{equation}
on the critical bubble solution for  a relativistic metastable fluid
given in \cite{ck} which has been used for the description of the
QCD phase transition. There the dynamical growth rate of the bubble
was described by taking into account the dissipative effects of a
fluid with an energy momentum tensor
\begin{equation}
T_{\mu\nu} = p\,g_{\mu\nu} + \,(\rho +p) u_{\mu}\, u_{\nu} +
\tilde{T}_{\mu\nu}
\end{equation}
with
\begin{eqnarray}
\tilde{T}_{\mu\nu} = &-&\eta(\partial_{\mu}u_{\nu}
+\partial_{\nu}u_{\mu}+u_{\mu}u^{\alpha}\partial_{\alpha}u_{\nu}
+u_{\nu}u^{\alpha}\partial_{\alpha}u_{\mu}) \nonumber\\
&-&(\zeta-2\eta/3)(\partial_{\alpha}u^{\alpha})(g_{\mu\nu}+u_{\mu}u_{\nu}).
\end{eqnarray}
Here $\rho$ and $p$ are the fluid energy density and pressure
respectively and $u^{\mu}=(1,\vec{v})$ is the four-velocity for a
relativistic fluid. As an additional phenomenological input one
needs the shear and bulk viscocities, $\eta$ and $\zeta$
respectively.

The dynamical growth rate, $\kappa$, is calculated by considering a
spherically symmetric, exponentially growing, perturbation
\begin{equation}
\gamma(\vec{r}, t) = \gamma(r) \, e^{\kappa t},
\end{equation}
\begin{equation}
\vec{v}(\vec{r}, t) = v(r) \, e^{\kappa t},
\end{equation}
of the also spherically symmetric metastable configuration of a
critical bubble $\rho = \bar{\rho}(r), \, \vec{v}=0$. When $
\rho(\vec{r}, t) = \bar{\rho}(r) + \gamma(\vec{r}, t)$ and the
remaining equations are used in the equations of motion,
self-consistency of the solution determines $\kappa$.

As explained before, I will calculate the first gravitational
corrections by determining the effects of the Hubble drag term,
$H=\dot{R}/R$, on the results of \cite{ck}. Thus I will assume that
a natural free energy description of the problem exists in the rest
frame of the fluid. Although the considerations here are similar to
works on stability of classical and solitonic configurations in an
expanding universe \cite{cooperstock, copeland, metaxas}, it is
important to realize that we are not investigating quite the same
problem; we are rather interested in the effects of cosmological
expansion on the actual flat space-time growth rate, $\kappa_0$, of
the already unstable (metastable) solution.

The equations of motion
\begin{equation}
\nabla_{\mu} \, T^{\mu\nu} = 0
\end{equation}
can be solved approximately in two regions: region (I) which extends
from just inside the bubble radius $R_*$ to a few correlation
lengths, $\xi$, outside the bubble surface, and region (II) in
distances $r$ greater than the bubble radius $R_*$ plus a few
correlation lengths. The analysis of \cite{ck} shows that the fluid
velocity $v(r)$ has the following behavior: it is very close to zero
from the center of the bubble up to a few $\xi$s inside the bubble
surface, then it rises abruptly until in region (I) it starts to
fall like $1/r^2$ and in region (II) it falls exponentially to zero.
This behavior will be modified by the Hubble  drag term, and by
matching the two solutions at $r\approx R_* +$ a few $\xi$s, one
obtains the corrected $\kappa = \kappa_0 + \delta \kappa$.

The notation is as follows: overbars will denote the critical bubble
solution, subscripts $0$ will denote the flat space-time values,
subscripts I and II the respective regions and $\Delta$ will denote
the difference of a quantity between the equilibrium (subscript t)
and the metastable (subscript f) phase. Also, as before, $\xi$ will
denote the correlation length and $R_*$ the critical bubble radius
in the presence of gravity.

 One will also make use of
the enthalpy density, $w=\rho + p$, and the bubble surface tension,
$\sigma$, in terms of which the flat case growth rate has been
calculated as \cite{ck}
\begin{equation}
\kappa_0 = \frac{4\sigma(\zeta +4\eta/3)}{(\Delta w)^2 \,R_0^2}.
 \label{flat}
\end{equation}
and the kinematical prefactor as
\begin{equation}
\Omega_0 = \frac{2}{3\sqrt{3}}\left(\frac{\sigma}{T}\right)^{3/2}
\left(\frac{R_0}{\xi}\right)^4 \,Vol
\end{equation}
where the volume of the system, $Vol$, is usually divided out as in 
(\ref{basic}).

 One expects these relations to carry over in the presence 
of gravity, in our approximations, by replacing $R_0$ by $R_*$, keeping the 
surface tension $\sigma$ approximately the same in the thin wall approximation,
 and the
physical volume of the system giving a factor of $\sqrt{-g}$.
The cosmological expansion, however, described by the Hubble drag term,
will give an additional contribution to the dynamical prefactor, that can be 
calculated from the equations of motion.

 The $\nu =0$ equation of motion
\begin{equation}
\kappa\gamma(r)=-\frac{1}{r^2}\frac{d}{dr}[r^2 \, \bar{w}
v(r)]\,-3H\bar{w}
 \label{nu=0}
\end{equation}
gives in region (I)
\begin{equation}
v_{\rm I} (r)= \frac{C}{r^2} + Hr + H\frac{R_*^3}{r^2}\frac{\Delta
w}{w_f}
\end{equation}
and the $\nu=i$ equation of motion can be simplified in region (II)
as
\begin{equation}
(\kappa + 3 H) \bar{w} v(r) = (\zeta
+4\eta/3)\frac{d}{dr}\left(\frac{1}{r^2}\frac{d}{dr}[r^2
v(r)]\right)
 \label{nu=i}
\end{equation}
to give
\begin{equation}
v_{\rm II} (r) = D \left( \frac{\alpha}{r} + \frac{1}{r^2} \right)
e^{-\alpha r}
\end{equation}
where
\begin{equation}
\alpha^2 = \frac{(\kappa + 3H) w}{(\zeta + 4\eta/3)}.
\end{equation}

The solutions depend on two constants, $C$ and $D$, the matching is
done for definiteness at $r = R_* + c\,\xi$, with $c$ a numerical
constant of order unity, and we get another condition from the fact
that, for $H=0$, the solution should be consistent with the previous
result (\ref{flat}). The final result for the corrected $\kappa =
\kappa_0 +\delta\kappa$ is
\begin{equation}
\delta\kappa = \kappa_0 H \sqrt{\frac{R_*}{c\,\xi}}
               \sqrt{\frac{R_* (\Delta w)}{\sigma}}
               \left( R_* + \frac{1}{\alpha_0} \right)
               \left( 1+\frac{\Delta w}{w_f} \right)
\label{result}
\end{equation}

In order to get an order-of-magnitude estimate for this result one
can assume that for sufficiently high temperature $T$ one can
approximate $\eta, \zeta \sim T^3/\lambda^2$, $\sigma\sim
T^3/\lambda$ to finally obtain the estimate for the corrections to
the prefactor that arise from the dynamical growth rate:
\begin{equation}
A \sim \mu^3 \left( T + \frac{H}{\lambda^{3/2}} \right).
 \label{resultdim}
\end{equation}

We see that the leading effect of the cosmological expansion has
been, indeed, to increase the bubble nucleation rate, and can be
significant when $H\sim\lambda^{3/2} T$ while, at the same time,
$H<<\mu$, in accordance with our approximations.
The surface tension was assumed unchanged in the thin wall approximation,
the correction to the bubble radius can be estimated from 
(\ref{deltaR}), it is also subleading, however, in the approximation used.
 It should be
noted that this result will also be modified when gravitational
corrections to $\Omega$, the critical bubble radius,  and surface tension are
incorporated, when one goes beyond the limits of our approximations.
In any case, even when other physical situations are considered, the corrections
estimated here also exist and can be calculated, for example, by
a self-consistent solution of equations like (\ref{nu=0}) and (\ref{nu=i}).
 What is more important, and supportive of the
arguments of the previous section,however, is that the corrections estimated
here are different than what is usually assumed as a naive, dimensional
pre-exponential factor, for example $\mu^4$ or $T^4$, and have an
interesting dependence on $\Lambda$ (or $H$).

\section{Comments}

The main purpose of this work has been to motivate the suggestion
that a fuller treatment of the theory of metastability in curved
space-time is needed, in order to supply the Coleman-De~Luccia
result with possible additional gravitational corrections that may
provide valuable insight to applications in cosmological problems.

One way to approach this problem is to attempt a generalization of
Langer's original theory of statistical metastability. The main
difficulties of this approach stem from the fact that proper
definitions of the thermodynamical quantities are needed, presumably
with the use of relativistic kinetic theory in the expanding
universe. It was implicitly assumed here that such an extension can
be done in the fluid's rest frame and the first corrections to the
flat space-time results that were presented show, indeed, the
expected increase in the bubble nucleation rate due to the
cosmological expansion. The fuller treatment of the relativistic
thermodynamics of first-order phase transitions is expected to give
additional contributions to the nucleation rate, similar to the ones
presented here and generally different than the usually assumed
pre-exponential factor.

\vspace{0.5in}

\centerline{\bf Acknowledgements}

This work was completed while visiting the National Technical
University of Athens. I am grateful to the people of the Physics
Department for their hospitality.

\end{document}